\documentclass[12pt]{article}
\pdfoutput=1
\usepackage[yyyymmdd,hhmmss]{datetime}
\usepackage{hyperref}
\usepackage{cite}
\usepackage{color}
\usepackage{graphicx}
\usepackage{amsmath}
\usepackage{amssymb}
\usepackage{xspace}

\makeatletter
\@addtoreset{equation}{section}

\makeatletter
\renewcommand\section{\@startsection {section}{1}{\z@}%
                                   {-3.5ex \@plus -1ex \@minus -.2ex}
                                   {2.3ex \@plus.2ex}%
                                   {\normalfont\large\bfseries}}
\renewcommand\subsection{\@startsection{subsection}{2}{\z@}%
                                     {-3.25ex\@plus -1ex \@minus -.2ex}%
                                     {1.5ex \@plus .2ex}%
                                     {\normalfont\bfseries}}

\def\baselinestretch{1.2}
\newcommand{\be}{\begin{equation}}
\newcommand{\ee}{\end{equation}}
\newcommand{\beq}{\begin{eqnarray}}
\newcommand{\eeq}{\end{eqnarray}}

\newcommand{\gone}[1]{{}}

\begin{document}
\begin{titlepage}

\rule{0ex}{0ex}

\vfil

\begin{center}

{\bf \Large   Comments on  $T \bar T$ double trace deformations \\
  and boundary conditions
}

\vfil

William Cottrell$^{1,2}$ and   Akikazu Hashimoto$^3$

\vfil

{}$^1$ Stanford Institute for Theoretical Physics and Department of Physics \\ Stanford University, Stanford, CA 94305-4060, USA

{}$^2$ Institute for Theoretical Physics Amsterdam, University of Amsterdam\\ 1098 XH Amsterdam, The Netherlands

{}$^3$ Department of Physics, University of Wisconsin, Madison, WI 53706, USA

\vfil

\end{center}

\begin{abstract}
\noindent We study the UV dynamics of $\mu T \bar T$ deformed
conformal field theories formulated as a deformation of generating
functions. We explore the issue of non-perturbative completion of the
$\mu$ expansion by deriving an integral expression using the
Fourier/Legendre transform technique, and show that it is more natural
to impose Neumann, as opposed to the Dirichlet, boundary condition,
for the metric at the cut-off surface recently proposed by McGough,
Mezei, and Verlinde. We also comment on interesting connection to
boundary conformal field theories.
\end{abstract}
\vspace{0.5in}

\end{titlepage}
\renewcommand{\baselinestretch}{1.05}  

There is a very interesting proposal \cite{McGough:2016lol} that a CFT
in 1+1 dimensions which also admits a holographic description is
deformable by an irrelevant operator of the form
\be \mu \int d z \, d
\bar z \ T(z) \bar T(\bar z) \label{mudef} \ee
and that the resulting system is 1) ultraviolet complete and
dynamically well defined as a quantum system, and 2) exhibits a
physical cut-off in the number of degrees of freedom in the
ultraviolet. This proposal is based mostly on the work of
\cite{Smirnov:2016lqw}, and was was subjected to tests in
\cite{McGough:2016lol} by comparing the group velocity of small
fluctuations at finite temperature and the general features of the
spectrum of states of the system.

Situations where an irrelevant deformation of a CFT actually makes sense
do not occur generically, although they are not
strictly forbidden either. It it something for which we don't have much
intuition mostly due to lack of experience. Examples that have been
explored can be found in \cite{Cavaglia:2016oda}. The fact that the
ultraviolet is cut off implies that these are not conformal field theories in the
usual sense.

The goal of this note is to explore the ultraviolet dynamics of this
system. We will take a closer look at the cut-off from the both the
field theory and holographic point of view in the context of AdS/CFT
correspondence. Specifically, we will suggest that it is more natural
to impose Neumann, rather than Dirichlet, boundary condition for the
metric at the holographic boundary in order to match the prescription
of \cite{Smirnov:2016lqw}. We will also offer several comments which
follow as a consequence of this proposal. Interesting discussions
relevant to this topic can be found in
\cite{Shyam:2017znq,Krishnan:2016dgy,Dubovsky:2017cnj,Cardy:2018sdv}. Also, a
slightly different perspective on irrelevant deformation with a
different UV completion can be found in
\cite{Giveon:2017nie,Giveon:2017myj,Asrat:2017tzd}.

We will approach this problem by assuming that a solution to a CFT, in
$d$ dimensions, has been provided abstractly. Solving a CFT implies
providing a list of all operators and their correlation
functions. Information needed to encode all of that can be presented
in the form of a generating functional
\be Z_D[g^\infty_{ij}(x),A_i^\infty(x), \phi^\infty(x), \ldots]  \ee
where the list of fields contained in the argument of $Z_D[\ldots]$ is
in one to one correspondence with the list of operators. The
functional derivative with respect to these fields will then generate
the correlation function. The field $g_{ij}$ is special in that it
encodes the rigid geometry on which the CFT lives, and variation with
respect to small perturbation of $g_{ij}(x)$ corresponds to the
insertion of the stress energy tensor. $Z_D[\ldots]$ can sometimes be
computed using bootstrap methods.  Another approach is to invoke the
AdS/CFT correspondence, in which case, the fields in the argument of
$Z_D[\ldots]$ are realized as dynamical fields living in an $AdS_{d+1}$
bulk. $Z_D[\ldots]$ is then computed by computing the quantum gravity
partition function where the bulk fields are usually subjected to Dirichlet
boundary condition\footnote{Which is why there is a subscript $D$ in $Z_D$ and
a superscript $\infty$ in the arguments of $Z_D[\ldots]$} at the
boundary of $AdS_{d+1}$. Of course, at the moment, the full quantum
gravity partition function is beyond the scope of what we are able to
practically compute, but one can extract information about
$Z_D[\ldots]$ reliably in the classical gravity approximation. This amounts to working to leading order in a large
\be {\cal N} = {L^{d-1} \over  16 \pi G_N} \ee
expansion, where $L$ is the radius of $AdS_{d+1}$ and $G_N$ is the
Newton constant in $d+1$ dimensions.

In this abstract formalism, it is easy to formulate what one means by (\ref{mudef}). One simply considers the deformed generating functional\footnote{The sign of $\mu$ is chosen that the positive $\mu$ corresponds to negative $\tilde \mu$ in the convention of figure 1 of \cite{McGough:2016lol}. We are also following the convention of \cite{McGough:2016lol} where    $T \bar T = T_{\mu \nu} T^{\mu \nu}/8- (T^\mu{}_\mu)^2/16$.}
\be Z_{def}[g^\infty_{ij}(x),A_i^\infty(x), \phi^\infty(x), \ldots] = e^{-{\mu \over 2} \int d^d x {\delta \over \delta g_{ij}^\infty(x)} g^\infty_{jk} {\delta \over \delta g^\infty_{kl}(x)} g^\infty_{ki}} Z_D[g^\infty_{ij}(x),A_i^\infty(x), \phi^\infty(x), \ldots] \ .
\ee
This expression is easy to interpret as an expansion in $\mu$ which
gives rise to an expression with natural conformal perturbation theory
interpretation. Formally, this expression appears to define a
generating function $Z_{def}[\ldots]$ for the deformed theory. The
formal expression also highlights the one important subtlety, namely
whether the expression is well defined non-perturbatively in $\mu$. In
other words, does the expansion in $\mu$ admits an unambiguous
resummation? If so $Z_{def}[\ldots]$ is completely well defined and
ultraviolet complete.

In order to gain a feel for this question, it is useful to explore the
analogous issue when the operators being inserted are scalars or
vectors. One can, for instance, consider a deformation of the type
\be Z_{scalar}[g^\infty_{ij}(x),A_i^\infty(x), \phi^\infty(x), \ldots] = e^{-\mu \int d^d x {\delta \over \delta \phi^\infty (x)}  {\delta \over \delta \phi^\infty(x)}} Z_D[g^\infty_{ij}(x),A_i^\infty(x), \phi^\infty(x), \ldots] \ . \label{scalardef} 
\ee
This deformation is interpretable as an insertion of double trace
operator built out of the operator sourced by $\phi^\infty(x)$ in the
standard AdS/CFT terminology.

For this case, the issue of resummation in $\mu$ can be addressed systematically. The idea is to recognize that derivatives are like conjugate variable. One formally inserts a functional delta function
\be \delta(\phi(x)-\varphi(x)) =
\int [D J] e^{i \int d^dx\, J(x)(\phi^\infty(x) - \varphi^\infty(x))}  \ee
so that we can write
\beq \lefteqn{ Z_{scalar}[g^\infty_{ij}(x),A_i^\infty(x), \phi^\infty(x), \ldots]} \cr
&=& \int [D \varphi^\infty][D J] e^{i \int d^dx\, J(x)(\phi^\infty(x) - \varphi^\infty(x))} e^{-\mu \int d^d x {\delta \over \delta \varphi^\infty (x)}  {\delta \over \delta \varphi^\infty(x)}} Z_D[g^\infty_{ij}(x),A_i^\infty(x), \varphi^\infty(x), \ldots] \cr
&=& \int [D \varphi^\infty][D J] e^{i \int d^dx\, J(x)(\phi^\infty(x) - \varphi^\infty(x))} e^{\mu \int d^d x \, J(x)^2} Z_D[g^\infty_{ij}(x),A_i^\infty(x), \varphi^\infty(x), \ldots] \cr
&=& \int [D \varphi^\infty] e^{\int
  d^dx\, \left( {1 \over 4 \mu} \varphi^\infty(x)^2 - {1 \over 2 \mu}
  \phi^\infty(x) \varphi^\infty(x) + {1 \over 4 \mu}
  \phi^\infty(x)^2\right)} Z_D[g^\infty_{ij}(x),A_i^\infty(x),
  \varphi^\infty(x), \ldots] \eeq

This is an expression of the form \cite{Casper:2017gcw} 
\beq \lefteqn{ Z_{scalar}[g^\infty_{ij}(x),A_i^\infty(x),
    \phi^\infty(x), \ldots]} \cr &=& \int [D \varphi^\infty] e^{- \int
  d^dx\, {\cal N} \left( \alpha \varphi^\infty(x)^2 +\beta
  \phi^\infty(x) \varphi^\infty(x) + \gamma 
  \phi^\infty(x)^2\right)} Z_D[g^\infty_{ij}(x),A_i^\infty(x),
  \varphi^\infty(x), \ldots] \label{Zscalar}\eeq
and is interpretable as the relevant deformation of the Neumann theory
by a double trace operator constructed out of the operator associated
to the bulk field $\phi(x)$ satisfying the Neumann boundary
condition\footnote{So $\phi^\infty(x)$ is the coefficient of the
  subleading term in the expansion near the boundary.}
\cite{Klebanov:1999tb,Witten:2001ua}. The expression (\ref{Zscalar})
is simply a convolution of $Z_D[\ldots]$ by a Gaussian, and as such
appears to be a well defined expression. Typically, in the standard
Dirichlet prescription, the scalar $\phi$ is assigned a dimension $2
\Delta_+$ where
\be \Delta_\pm= {d \pm \sqrt{d^2 + 4 m^2} \over 2}\ ,  \ee
and so the double trace operator built out of it is irrelevant.  What
we seem to have here is that in attempting to understand the
deformation (\ref{scalardef}) at the non-perturbative level, we are
naturally led to the full RG flow \cite{Gubser:2002vv,Hartman:2006dy},
which consists of a Neumann theory in the UV being deformed by a
relevant operator of dimension $2 \Delta_-$ and ultimately flowing to
the Dirichlet in the IR theory whose leading irrelevant deformation is
the dimension the $2 \Delta_+$ operator. We can also relate the
magnitude of the deformations from the UV and the IR perspectives of
this flow as follows:
\be \mu =- {1 \over  4 {\cal N} \alpha} \label{alphamu} \ . \ee

It is straight forward to generalize this analysis to the deformation by double trace operator of the Thirring type
\be \mu \int d^d x \,   J^\mu(x) J_\mu(x)\label{thirring} \ee
by considering
\be Z_{vector}[g^\infty_{ij}(x),A_i^\infty(x), \phi^\infty(x), \ldots] = e^{-\mu \int d^d x g_{ij}^\infty(x) {\delta \over \delta A_i^\infty (x)}  {\delta \over \delta A_j^\infty(x)}} Z_D[g^\infty_{ij}(x),A_i^\infty(x), \phi^\infty(x), \ldots] \ . \label{vectordef} 
\ee

Repeating the same steps which we took for the scalars, we arrive at
\beq \lefteqn{ Z_{vector}[g^\infty_{ij}(x),A_i^\infty(x),
    \phi^\infty(x), \ldots]} \cr &=& \int {[D a_i^\infty][D \sigma]  \over \mbox{Vol}(G)} e^{ \int
  d^dx\, {\cal N} \left( {1 \over 4 \mu} g^\infty_{ij} (a_i - \partial_i \sigma) (a_j - \partial_j \sigma) -{1 \over 2 \mu} (a_i - \partial_i \sigma) A^i +  {1 \over 4 \mu} A_i A^i  \right)}\cr
&& \qquad \qquad \qquad \qquad \qquad \qquad\qquad \qquad  \times Z_D[g^\infty_{ij}(x),a_i^\infty(x),
  \phi^\infty(x), \ldots] \ . \label{Zvector}\eeq
Precisely the same expression in slightly different notation was
presented in (4.5) of \cite{Cottrell:2017gkb}. The interpretation of
(\ref{Zvector}) is essentially identical to that of the scalars. In
particular, the Thirring deformation of the Dirichlet theory is being
interpreted as the IR limit of a Neumann theory in the UV with
mixed boundary conditions \cite{Marolf:2006nd}. The UV theory is once
again a Gaussian convolution of the IR theory.  The relation (\ref{alphamu}) between $\alpha$ and $\mu$ continues to hold. The only subtlety we
wish to highlight is that because the Gaussian convolution involves
integrating over a spin 1 field, care is needed to ensure that the
path integral can be set up in a manifestly gauge invariant form. We
have therefore made the quotient by the gauge orbit and the
Stueckelburg field manifest in (\ref{Zvector}).

We are now in a position to generalize these results to stress energy
tensor. To the extent that we are exploring Lengendre/Fourier
transform to convert the functions of functional derivatives to
functions of conjugate momenta, we should anticipate the analogues of
(\ref{Zscalar}) and (\ref{Zvector}) to involve functional integral
over the metric field $g_{ij}$. In other words, the resummation
requires quantum gravity. The problem of quantum gravity of course is
not solved in general. However, quantum gravity in $d=2$ is an
exceptional case where we do know how to perform the path integral
\cite{Polyakov:1981rd}. It appears that the reason the $T \bar T$
deformation of \cite{Smirnov:2016lqw} is UV complete is closely
related to the fact that quantum gravity in $d=2$ make sense.

Following a similar line of reasoning, we first define the Neumann theory for the tensor theory as follows
\be Z_N = \int {[Dg_{ij}^\infty] \over \mbox{Vol(Diff)}} Z_D[g_{ij}^\infty(x),a_i^\infty(x),\phi^\infty(x)] \ . \ee
Note that we have removed all arguments in $Z_N$. The reason is that
in a theory of gravity, we do not except there to be any gauge
invariant local operators with which to compute correlation functions
\cite{Weinberg:1980kq}. The observables that we are allowed to
consider are generally of the non-local type, and include world sheet
partition functions in various topologies as well as $S$-matrix
elements for the fluctuations on the world sheet.

In order to incorporate the relevant deformation of the UV theory to
mimic the RG flow to the IR theory, we need the analogue of the
$\alpha$ term but it needs to be presented in a form which is
manifestly invariant with respect to diffeomorphism. The natural candidate which behaves as a mass term for small fluctuations yet respect diffeomorphism invariance at the non-linear level is the bare cosmological constant term. We therefore have
\be Z_{tensor}[\alpha] = \int {[Dg_{ij}^\infty] \over
  \mbox{Vol(Diff)}} e^{-2 {\cal N} \alpha \int d^d x \sqrt{g_\infty}}
Z_D[g_{ij}^\infty(x),a_i^\infty(x),\phi^\infty(x)] \ . \label{Ztensor}
\ee
where we have made the $\alpha$ dependence of the partition function
manifest. Note also that bare cosmological constant is typically
included when considering the quantization of generic matter in $d=2$
coupled to quantum gravity in a Liouville formalism, reviewed e.g.\ in
\cite{Ginsparg:1993is}.

Our claim is that this ``theory'' can be interpreted as the UV
complete description of (\ref{mudef}) when $d=2$. One can define
correlation functions of local operators order by order as an expansion
in small $\mu$ and compare against predictions from conformal
perturbation theory. However, our claim based on the structure of
(\ref{Ztensor}) is that a concept of local observables do not exist
microscopically, and as such these observables are not well defined
non-perturbatively in $\mu$.

Let us comment on the holographic interpretation of this picture when
the CFT admits a good gravity description. In \cite{McGough:2016lol},
it was suggested that the UV cut-off wall be subjected to Dirichlet
boundary condition so that a notion of quasi-local observables can be
defined. We are suggesting instead that it is more natural to assign a
Neumann boundary condition. A physical setup where the AdS geometry is
subjected to an ``artificial'' boundary subject to Neumann boundary
condition was considered in a very different context by Karch and
Randall in \cite{Karch:2000ct}. The setup of Karch and Randall also
involves the bare cosmological constant, which they parameterize using
the variable
\be \lambda= 2 \alpha L \ . \ee
One of the main points of \cite{Karch:2000ct} is the fact that in
order for the geometry of the boundary to be flat (or equivalently, in
order to tune the effective cosmological constant to zero), the bare
cosmological constant needed to be tuned to a finite value. Their
analysis for the case of $d=4$ can be found in (3) of
\cite{Karch:2000ct}. In our context, this implies that tuning the RG
flow to go into a CFT on flat space, the $\alpha$ must be
non-vanishing. Consequently,
\be {\mu \over 2}  = -{1 \over  4 {\cal N} \alpha} \ee
also needs to be non-vanishing.\footnote{This also suggests that limits  $\mu
  \rightarrow 0_+$ and $\mu \rightarrow 0_-$ may be disconnected.}
From the perspective of \cite{Karch:2000ct}, we can also relate the
value of $\lambda = (d-1) r_c^2/L^3$ to the position of the cut-off
surface\footnote{The value of $\lambda$ reported by \cite{Karch:2000ct} was for $d=3$ and in a conformal frame where $r_{c}=L$.  One can easily generalize via a conformal transformation.}  in Poincare coordinate, and ${\cal N} = c/24 \pi$
\cite{Brown:1986nw}. In this way, we essentially recover the same
relation\footnote{Recall that our conventions for $\mu$ are the opposite of \cite{McGough:2016lol}.}
\be \mu = -{24 \pi L^4 \over c r_c^2} \ee
as (1.3) of \cite{McGough:2016lol}, although we are
stressing that the boundary condition on the cut-off surface that we
are imposing is Neumann. This has important consequences on the set of
physical observables.

We also note that taking $\alpha$ to
be smaller than the critical value necessary for flat space, then, according to \cite{Karch:2000ct}, we would generate a
negative effective cosmological constant. This appears then to lead
naturally to the setup considered in \cite{Dubovsky:2017cnj}. 

The appearance of Neumann boundary also offers an interesting
connection to the construction of boundary conformal field theory 
\cite{Takayanagi:2011zk}. In order to make this connection a bit more
tangible, it is useful to tune the effective cosmological constant so
that the cut-off boundary has the geometry of $AdS_2$. In that case,
the cut-off boundary is precisely identifiable as the $Q$ component of
the boundary using the notation of \cite{Takayanagi:2011zk}.  We refer
the reader to figure 1 of \cite{Takayanagi:2011zk} where the $Q$ and
$M$ components of the boundary of a bulk AdS geometry are
defined. The same figure is reproduced in figure \ref{figb} below. The
observables of BCFT are local operators inserted in the $M$
component. These are interpretable as ``boundary observables'' living
on the boundary of the ``boundary'' $AdS_2$. We however do not insert
operators on the Neumann surface $Q$. One can then imagine taking the
limit of vanishing effective cosmological constant on $AdS_2$. This
amounts to taking the flat space limit where the boundary observables
on $AdS_2$ becomes the $S$-matrix on flat space, along the lines of
\cite{Polchinski:1999ry}. From this perspective, it is clear that
there are non-trivial observables such as the $S$-matrix elements one
can compute even when the notion of local correlation functions are
absent. At the same time, we also see that identifying correct set of
observables are more subtle when quantum gravity is involved.

\begin{figure}
  \centerline{\includegraphics{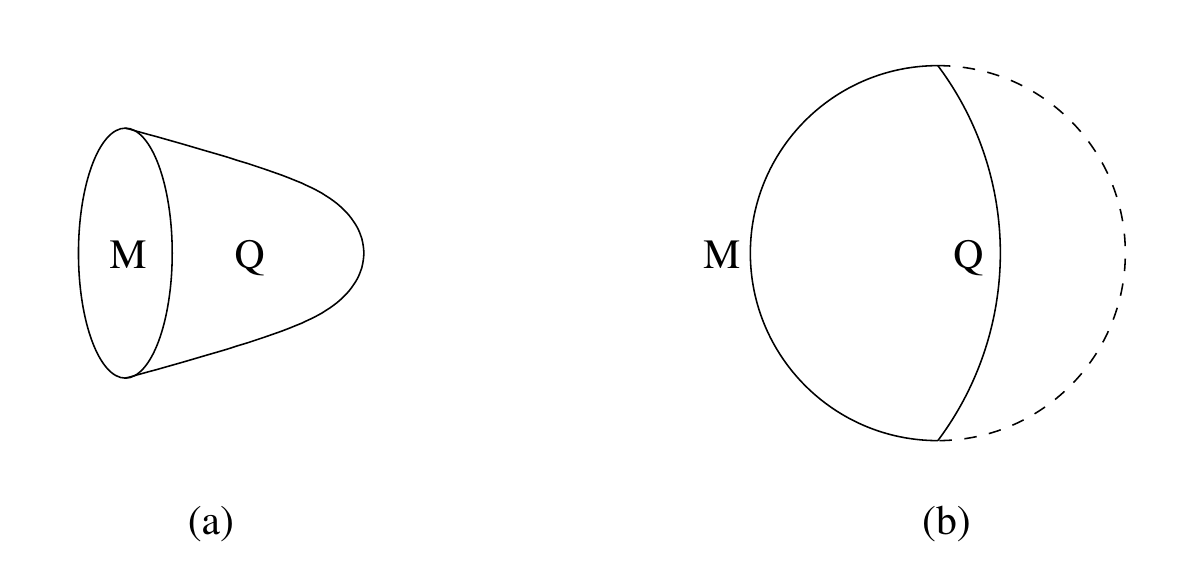}}
  \caption{Illustration of locally $AdS_{d+1}$ bulk  spacetime $N$ bounded by boundary components $M$ and $Q$. (a) is a reproduction of figure 1 of \cite{Takayanagi:2011zk}. (b) is the embedding of (a) inside global anti de Sitter geometry as was illustrated in figure 6 of \cite{Karch:2000ct}. The spacetime of boundary field theory from the $T \bar T$ deformed CFT point of view is $Q$. Here, we are taking the effective cosmological constant to be negative so that the geometry on $Q$ is $AdS_d$. From the boundary field theory point of view, $M$ is an auxiliary structure which emerges as a dual representation of large number of degrees of freedom that the original CFT contained. Operators are inserted on $M$ and are to be interpreted as boundary observables for the gravity theory on $AdS_d$, or as $S$-matrix elements in the limit where $AdS_d$ approaches flat $d$-dimensional space-time.    \label{figb}}
\end{figure}  

Finally, let us offer few pieces of evidence in support of our claim that the UV complete description of $T \bar T$ deformed CFT in $d=2$ must be of
Neumann type.

One argument is the fact that all observables computed in the
literature (to the best of our knowledge) are consistent
with the Neumann interpretation. The partition function on a cylinder
or a torus, and the $S$-matrix elements, are typically reported to
establish the theory being well defined at the quantum level
\cite{Smirnov:2016lqw,Cavaglia:2016oda}. Local correlation functions
have been computed in \cite{Kraus:2018xrn} but only to few orders as
an expansion in $\mu$. The issue of whether this series is summable at
the non-perturbative level appears to be left open in most of these
discussions. Here, we have provided an argument based on the structure
of Fourier/Legendre transform and gauge invariance.  Related arguments have also recently been provided by \cite{Cardy:2018sdv}.

\begin{figure}
  \centerline{\includegraphics{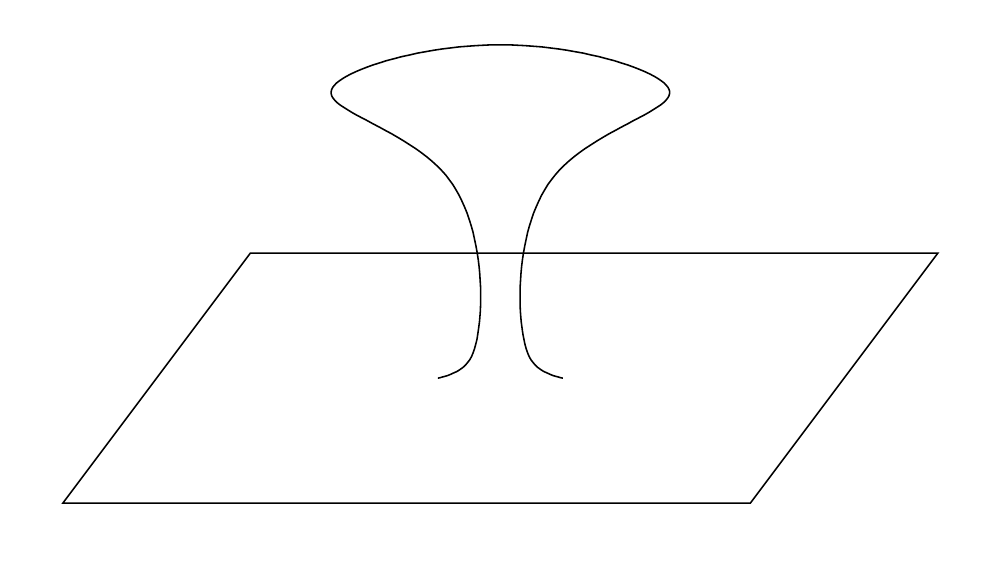}}
  \caption{An example of non-single valued field configuration which are allowed in Nambu-Goto theory. Similar issue arises in the ``tree stump'' configuration of BIons illustrated in figure 2 of  \cite{Callan:1997kz}.  \label{figa}}
\end{figure}

Another argument that quantum gravity must be involved in the full story
can be made by building on the observation that the $T \bar T$ deformation
can formally be resummed for the case of free scalar fields to bring
it into the form of Nambu-Goto action in static gauge
\cite{Cavaglia:2016oda}. The question is whether this theory should be
thought of as a gauge fixed version of a gauge invariant theory. If
so, the full quantization of the Nambu-Goto action must essentially
consist of performing the Polyakov path integral
\cite{Polyakov:1981rd}. It is difficult to imagine how to quantize the
Nambu-Goto action without invoking Polyakov's technique. One of the
salient features of Nambu-Goto and DBI action \cite{Callan:1997kz} is
the fact that a configuration of the type illustrated in figure
\ref{figa} where the embedding is not single valued are included in
the space of allowed configurations, and in fact has finite action. In
other words, singularities corresponding to the derivative of the
embedding field reaching infinity is coordinate, as opposed to
physical singularity. The issue of how to precisely identify the
singular configurations which should, or should not, be included in the path integral, is in fact the central challenge in properly understanding the gauge
field formulation of quantum gravity in $d=3$
\cite{Witten:1988hc,Witten:2007kt}. In $d=2$, by invoking diffeomorphism invariance and the uniformization theorem, one can iron out the non-singlevaluedness of configurations like the one illustrated in figure \ref{figa}, e.g.\ by going to lightcone gauge \cite{GODDARD1973109,Dubovsky:2012wk, Caselle:2013dra,Dubovsky:2015zey}.

Polyakov's path integral \cite{Polyakov:1981rd} is an elegant solution
which overcomes all of these subtleties. It is certainly possible, but
it is difficult for us to imagine an alternative, consistent,
quantization of a long Nambu-Goto string aside from the Polyakov's
prescription. At the end of section 4.1 of \cite{McGough:2016lol}, the
authors indeed state that the $T \bar T$ system ``behaves like a
causal theory similar to 2D quantum gravity.'' We are making a
stornger statement that {\it is} a theory of 2D quantum gravity, in
the sense that it is a quantum theory with diffeomorphism invariance.

That the boundary condition for the metric is Neumann instead of
Dirichlet will only affect the UV of the full system. Most of the
effective physics in the IR are unaffected. Nonetheless, since the $T
\bar T$ deformation of \cite{Smirnov:2016lqw} is supposed to be UV
complete, the distinction between Neumann and Dirichlet is physically
meaningful.

It should also be re-iterated that the functional Fourier/Legendre
transform made sense mathematically without additional input only for
$d=2$. This suggests that the generalization of \cite{Smirnov:2016lqw}
will likely not work so nicely in dimensions other than $d=2$. It
would be very interesting to explore if the special features of
quantum gravity in $d=3$ can be exploited to push the agenda in that
direction \cite{Witten:1988hc,Witten:2007kt}.  Perhaps some version of the story in higher dimension can be constructed following the prescription of \cite{Verlinde:1999fy}.

Finally, let us comment in closing that while we have provided
arguments based on the structure of Fourier/Legendre transform and
some circumstantial observations, we have not provided a rigorous
argument that conformal perturbation theory in $\mu$ in fact fails to
converge. It is likely that correlation functions suffer from
breakdown of locality at short distances along the lines found in
\cite{Chen:2017dnl}.  It would be very interesting to understand the nature of the asymptotic expansion in $\mu$ more thoroughly.  

\section*{Acknowledgements}

This work is supported in part by the DOE grant DE-SC0017647.  We would also like to thank Victor Gorbenko and Mehrdad Mirbabayi for interesting conversations.

\bibliography{bcft}\bibliographystyle{utphys}

\providecommand{\href}[2]{#2}\begingroup\raggedright\begin{thebibliography}{10}

\bibitem{McGough:2016lol}
L.~McGough, M.~Mezei, and H.~Verlinde, ``{Moving the CFT into the bulk with
  $T\bar T$},''
\href{http://www.arXiv.org/abs/1611.03470}{{\tt 1611.03470}}.

\bibitem{Smirnov:2016lqw}
F.~A. Smirnov and A.~B. Zamolodchikov, ``{On space of integrable quantum field
  theories},'' {\em Nucl. Phys.} {\bf B915} (2017) 363--383,
\href{http://www.arXiv.org/abs/1608.05499}{{\tt 1608.05499}}.

\bibitem{Cavaglia:2016oda}
A.~Cavaglià, S.~Negro, I.~M. Szécsényi, and R.~Tateo, ``{$T
  \bar{T}$-deformed 2D Quantum Field Theories},'' {\em JHEP} {\bf 10} (2016)
  112,
\href{http://www.arXiv.org/abs/1608.05534}{{\tt 1608.05534}}.

\bibitem{Shyam:2017znq}
V.~Shyam, ``{Background independent holographic dual to $T\bar{T}$ deformed CFT
  with large central charge in 2 dimensions},'' {\em JHEP} {\bf 10} (2017) 108,
\href{http://www.arXiv.org/abs/1707.08118}{{\tt 1707.08118}}.

\bibitem{Krishnan:2016dgy}
C.~Krishnan, A.~Raju, and P.~N.~B. Subramanian, ``{Dynamical boundary for anti
  €"de Sitter space},'' {\em Phys. Rev.} {\bf D94} (2016), no.~12, 126011,
\href{http://www.arXiv.org/abs/1609.06300}{{\tt 1609.06300}}.

\bibitem{Dubovsky:2017cnj}
S.~Dubovsky, V.~Gorbenko, and M.~Mirbabayi, ``{Asymptotic fragility, near
  AdS$_{2}$ holography and $ T\overline{T} $},'' {\em JHEP} {\bf 09} (2017)
  136,
\href{http://www.arXiv.org/abs/1706.06604}{{\tt 1706.06604}}.

\bibitem{Cardy:2018sdv}
J.~Cardy, ``{The $T\overline T$ deformation of quantum field theory as a
  stochastic process},''
\href{http://www.arXiv.org/abs/1801.06895}{{\tt 1801.06895}}.

\bibitem{Giveon:2017nie}
A.~Giveon, N.~Itzhaki, and D.~Kutasov, ``{$T \bar T$ and LST},'' {\em JHEP}
  {\bf 07} (2017) 122,
\href{http://www.arXiv.org/abs/1701.05576}{{\tt 1701.05576}}.

\bibitem{Giveon:2017myj}
A.~Giveon, N.~Itzhaki, and D.~Kutasov, ``{A solvable irrelevant deformation of
  $AdS_3/CFT_2$},''
\href{http://www.arXiv.org/abs/1707.05800}{{\tt 1707.05800}}.

\bibitem{Asrat:2017tzd}
M.~Asrat, A.~Giveon, N.~Itzhaki, and D.~Kutasov, ``{Holography beyond $AdS$},''
\href{http://www.arXiv.org/abs/1711.02690}{{\tt 1711.02690}}.

\bibitem{Casper:2017gcw}
S.~Casper, W.~Cottrell, A.~Hashimoto, A.~Loveridge, and D.~Pettengill,
  ``{Stability and boundedness in AdS/CFT with double trace deformations},''
\href{http://www.arXiv.org/abs/1709.00445}{{\tt 1709.00445}}.

\bibitem{Klebanov:1999tb}
I.~R. Klebanov and E.~Witten, ``{AdS/CFT correspondence and symmetry
  breaking},'' {\em Nucl. Phys.} {\bf B556} (1999) 89--114,
\href{http://www.arXiv.org/abs/hep-th/9905104}{{\tt hep-th/9905104}}.

\bibitem{Witten:2001ua}
E.~Witten, ``{Multitrace operators, boundary conditions, and AdS/CFT
  correspondence},''
\href{http://www.arXiv.org/abs/hep-th/0112258}{{\tt hep-th/0112258}}.

\bibitem{Gubser:2002vv}
S.~S. Gubser and I.~R. Klebanov, ``{A universal result on central charges in
  the presence of double trace deformations},'' {\em Nucl. Phys.} {\bf B656}
  (2003) 23--36,
\href{http://www.arXiv.org/abs/hep-th/0212138}{{\tt hep-th/0212138}}.

\bibitem{Hartman:2006dy}
T.~Hartman and L.~Rastelli, ``{Double-trace deformations, mixed boundary
  conditions and functional determinants in AdS/CFT},'' {\em JHEP} {\bf 01}
  (2008) 019,
\href{http://www.arXiv.org/abs/hep-th/0602106}{{\tt hep-th/0602106}}.

\bibitem{Cottrell:2017gkb}
W.~Cottrell, A.~Hashimoto, A.~Loveridge, and D.~Pettengill, ``{Stability and
  boundedness in AdS/CFT with double trace deformations II: Vector Fields},''
\href{http://www.arXiv.org/abs/1711.01257}{{\tt 1711.01257}}.

\bibitem{Marolf:2006nd}
D.~Marolf and S.~F. Ross, ``{Boundary conditions and new dualities: Vector
  fields in AdS/CFT},'' {\em JHEP} {\bf 11} (2006) 085,
\href{http://www.arXiv.org/abs/hep-th/0606113}{{\tt hep-th/0606113}}.

\bibitem{Polyakov:1981rd}
A.~M. Polyakov, ``{Quantum geometry of bosonic strings},'' {\em Phys. Lett.}
  {\bf 103B} (1981)
207--210.

\bibitem{Weinberg:1980kq}
S.~Weinberg and E.~Witten, ``{Limits on massless particles},'' {\em Phys.
  Lett.} {\bf 96B} (1980)
59--62.

\bibitem{Ginsparg:1993is}
P.~H. Ginsparg and G.~W. Moore, ``{Lectures on 2-D gravity and 2-D string
  theory},'' in {\em {Theoretical Advanced Study Institute (TASI 92): From
  Black Holes and Strings to Particles Boulder, Colorado, June 3-28, 1992}},
  pp.~277--469.
\newblock 1993.
\newblock
\href{http://www.arXiv.org/abs/hep-th/9304011}{{\tt hep-th/9304011}}.
\newblock

\bibitem{Karch:2000ct}
A.~Karch and L.~Randall, ``{Locally localized gravity},'' {\em JHEP} {\bf 05}
  (2001) 008,
\href{http://www.arXiv.org/abs/hep-th/0011156}{{\tt hep-th/0011156}}.

\bibitem{Brown:1986nw}
J.~D. Brown and M.~Henneaux, ``{Central charges in the canonical realization of
  asymptotic symmetries: An example from three-dimensional gravity},'' {\em
  Commun. Math. Phys.} {\bf 104} (1986)
207--226.

\bibitem{Takayanagi:2011zk}
T.~Takayanagi, ``{Holographic dual of BCFT},'' {\em Phys. Rev. Lett.} {\bf 107}
  (2011) 101602,
\href{http://www.arXiv.org/abs/1105.5165}{{\tt 1105.5165}}.

\bibitem{Polchinski:1999ry}
J.~Polchinski, ``{$S$-matrices from AdS space-time},''
\href{http://www.arXiv.org/abs/hep-th/9901076}{{\tt hep-th/9901076}}.

\bibitem{Kraus:2018xrn}
P.~Kraus, J.~Liu, and D.~Marolf, ``{Cutoff AdS$_3$ versus the $T\bar{T}$
  deformation},''
\href{http://www.arXiv.org/abs/1801.02714}{{\tt 1801.02714}}.

\bibitem{Callan:1997kz}
C.~G. Callan and J.~M. Maldacena, ``{Brane dynamics from the Born-Infeld
  action},'' {\em Nucl. Phys.} {\bf B513} (1998) 198--212,
\href{http://www.arXiv.org/abs/hep-th/9708147}{{\tt hep-th/9708147}}.

\bibitem{Witten:1988hc}
E.~Witten, ``{(2+1)-dimensional gravity as an exactly soluble system},'' {\em
  Nucl. Phys.} {\bf B311} (1988)
46.

\bibitem{Witten:2007kt}
E.~Witten, ``{Three-dimensional gravity revisited},''
\href{http://www.arXiv.org/abs/0706.3359}{{\tt 0706.3359}}.

\bibitem{GODDARD1973109}
P.~Goddard, J.~Goldstone, C.~Rebbi, and C.~Thorn, ``Quantum dynamics of a
  massless relativistic string,'' {\em Nucl. Phys.} {\bf B56} (1973), no.~1,
  109 -- 135.

\bibitem{Dubovsky:2012wk}
S.~Dubovsky, R.~Flauger, and V.~Gorbenko, ``{Solving the simplest theory of
  quantum gravity},'' {\em JHEP} {\bf 09} (2012) 133,
\href{http://www.arXiv.org/abs/1205.6805}{{\tt 1205.6805}}.

\bibitem{Caselle:2013dra}
M.~Caselle, D.~Fioravanti, F.~Gliozzi, and R.~Tateo, ``{Quantisation of the
  effective string with TBA},'' {\em JHEP} {\bf 07} (2013) 071,
\href{http://www.arXiv.org/abs/1305.1278}{{\tt 1305.1278}}.

\bibitem{Dubovsky:2015zey}
S.~Dubovsky and V.~Gorbenko, ``{Towards a theory of the QCD string},'' {\em
  JHEP} {\bf 02} (2016) 022,
\href{http://www.arXiv.org/abs/1511.01908}{{\tt 1511.01908}}.

\bibitem{Verlinde:1999fy}
H.~L. Verlinde, ``{Holography and compactification},'' {\em Nucl. Phys.} {\bf
  B580} (2000) 264--274,
\href{http://www.arXiv.org/abs/hep-th/9906182}{{\tt hep-th/9906182}}.

\bibitem{Chen:2017dnl}
H.~Chen, A.~L. Fitzpatrick, J.~Kaplan, and D.~Li, ``{The AdS$_3$ propagator and
  the fate of locality},''
\href{http://www.arXiv.org/abs/1712.02351}{{\tt 1712.02351}}.

\end{thebibliography}\endgroup

\end{document}